\documentclass[twocolumn,amsmath,amssymb,prl]{revtex4}

\usepackage{graphicx}
\usepackage{dcolumn}
\usepackage{bm}

\begin{document}

\title{Statistical Properties of Demand Fluctuation in the Financial Market}


\author{Kaushik~Matia$^1$, Kazuko~Yamasaki$^{1,2}$ }
\affiliation{ $^1$Center for Polymer Studies and Department of Physics, Boston
University, Boston, MA 02215 USA.\\
$^2$Tokyo University of Information Sciences, Chiba city 265-8501 Japan.\\
}


\date{working paper last revised: \today}

\begin{abstract}

We examine the out-of-equilibrium phase reported by Plerou {\it et. al.}
in Nature, {\bf 421}, 130 (2003) using the data of the New York stock
market (NYSE) between the years 2001 --2002. We find that the observed
two phase phenomenon is an artifact of the definition of the control
parameter coupled with the nature of the probability distribution
function of the share volume. We reproduce the two phase behavior by a
simple simulation demonstrating the absence of any collective
phenomenon. We further report some interesting statistical regularities
of the demand fluctuation of the financial market.

\end{abstract}
%
%

\maketitle

Recently a report based on New York stock exchange (NYSE) data between
period 1995-1996 ref.~\cite{Stanley1} reported that the financial market
has two phases, namely, the ``equilibrium'' phase, and the
``out-of-equilibrium'' phase.  Ref.~\cite{Stanley1} further reported a
critical point which is the boundary between a $0$ and finite value of
the order parameter for the observed phase transition.

In this paper we address the following questions in an effort to
understand the observed two phase phenomenon:
\begin{enumerate}
\item What is the cause of the observed two phase behavior ?
\item Do large changes of price occur in out-of-equilibrium phase ?
\end{enumerate}
Our study is based on the NYSE Trades and Quotes (TAQ) database for the
period 2001--2002 which records every ask, bid, and transaction price.

\begin{figure}
\begin{center}
\includegraphics[width=0.32\textwidth,angle=-90]{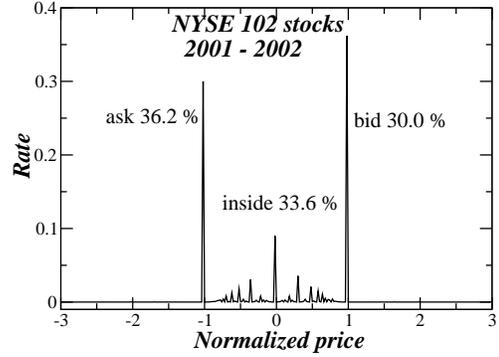}
\end{center}
\caption{ Percentage of transactions occurring at ask, bid and/or an
intermediate price between ask and bid prices which are normalized to +1
and -1, respectively. We observe that nearly 1/3 of the transactions
take place in each one of these three groups. }
\label{askbid1}
\end{figure}

Figure~\ref{askbid1} gives an estimate of the percentage of transactions
during the period period 2001--2002 that was either at the ask price
(scaled to be 1), at the bid price (scaled to be -1), or at an
intermediate value between the ask and the bid prices.

Using the method of ref.~\cite{Stanley1} we first identify a buyer
initiated or a seller initiated transaction by a quantity $a$ which is
defined as follows:
\begin{equation}
a = \left\{ \begin{array}{l}
\hspace{0.2cm} 1 \hspace{0.5cm} \mathrm{if}~~~ p > (ask - bid)/2 \\
\hspace{0.2cm} 0 \hspace{0.5cm} \mathrm{if}~~~ p = (ask - bid)/2 \\ 
-1 \hspace{0.5cm} \mathrm{if}~~~ p < (ask - bid)/2 \\ 
 \end{array} \right.
\end{equation}
where $p$ is a transaction price. Usually a transaction is executed when
a new quote hits the lowest ask price or the highest bid price. 

As in ref.~\cite{Stanley1}, we define the volume imbalance $\Omega$ and
its standard deviation as
\begin{equation}
\Omega (t) = \sum\limits_{i = 1}^{N(t)} {q_i a_i } 
\end{equation}
\begin{equation}
\Sigma (t) =  < |q_i a_i  - < q_i a_i >_t | >_t 
\end{equation}
where $q_i$ is the share volume of the $i$th transaction within the period
$\Delta t$.

\begin{figure}
\begin{center}
\includegraphics[width=0.32\textwidth,angle=-90]{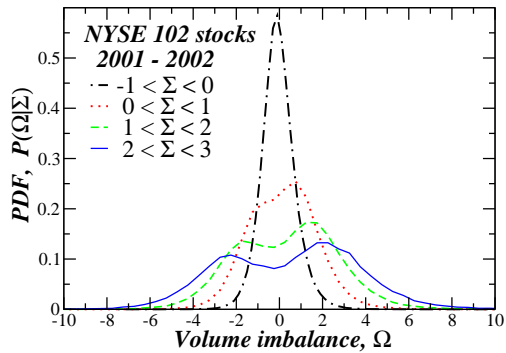}
\end{center}
\caption{Probability density function $P(\Omega|\Sigma)$ for given
$\Sigma$. We observe that P($\Omega|\Sigma$) for $\Sigma>
0$. Reference~\cite{Stanley1} interprets the two maximas of
P($\Omega|\Sigma$) as the two phases of the financial market. }
\label{askbid2}
\end{figure}

Fig.~\ref{askbid2} displays the probability density function (PDF)
$P(\Omega|\Sigma)$ of $\Omega$ for a given $\Sigma$. We observe a
bi-modal PDF with two extrema when $\Sigma$ is larger than a critical
value. This feature of $P(\Omega|\Sigma)$ is reported as an instance of
two phase behavior in the financial market [Fig. 1 (a)of
ref.~\cite{Stanley1}].

We next accumulate the $a$ between intervals $t$ and $t+\Delta t$ to
estimate a quantity $\Phi = \sum\limits_{i = 1}^{N(t)} {a_i} $ which
measures the number imbalance between buyer and seller initiated trade within
the time interval $\Delta t$.

The first moment $\sigma$ of $\Phi$ is defined as
\begin{equation}
\sigma (t) =  < |a_i  - < a_i >_t | >_t 
\end{equation}
where $<...>_t$ represents an average in the time interval $\Delta t$ .

We also define 
\begin{equation}
q(t) = \frac{1}{N(t)}\sum_{i=1}^{N(t)} q_i
\end{equation}
which measures the average volume of trade per transactions between $t$
and $t+\Delta t$.

We estimate $~\Omega (t),~\Sigma (t)$, and $q(t)$ averaged over 102
stocks which has the largest total volume among all NYSE stocks in 2002
with $\Delta t=15$ minutes. 
\footnote{ Because the second moments of $\Omega$ and $\Sigma$ are
expected to diverge, they were scaled in order that first moments, $<
|\Omega(t) - < \Omega(t) > | >$ and $< |\Sigma(t) - < \Sigma(t) > | >$
are equal to 1.  Where, $<...>$ means an average over whole period per
each stock.}.

In phase transitions observed in physical systems, the control
parameters are independent variables. But in ref.~\cite{Stanley1}, the
control parameter $\Sigma$ is the magnitude of fluctuation of the order
parameter $\Omega$. When fluctuations are large, the underlying PDF of
the random variable is wide, leading us to think that we might replicate
$P(\Omega|\Sigma)$ by simulation.

First we evaluate the empirical correlations present in the database
between different variables defined in eq.~1--5. Table~\ref{corr}
tabulates the estimated correlations.

\begin{table}
\begin{center}
\caption{Correlations}
\label{corr}
\begin{tabular}{|c|c|c|c|c|c|}
\hline
          & $|R|$ & $N$    & $q$  & $\Sigma$ & $\sigma$\\
\hline
$|R|$& 1.00       & 0.10   & 0.20 & 0.11     & 0.11\\
$N$       &   -        & 1.00   & 0.03 & 0.01     & 0.50\\
$q$       &   -        &   -    & 1.00 & 0.69     & 0.14\\
$\Sigma$  &   -        &   -    &   -  & 1.00     & 0.08\\
$\sigma$  &   -        &   -    &   -  &   -      & 1.00\\
\hline     
\end{tabular}
\end{center}
\end{table}

Next, for simulation, we consider $a$, $q$, and $N$ to be identically
and independently distributed (i.i.d) from two possible PDFs given in
Table~\ref{comb}~\cite{Stanley2}. We execute the simulation for eight
($2^3$) possible combinations of the set $a$, $q$ and $N$.

\begin{table}
\begin{center}
\caption{Different combinations of $a$, $q$ and $N$ for simulation. }
\label{comb}
\begin{tabular}{|c|c|c|}
\hline
$a$  &  random 1,-1   & Normal(0,1)\\
\hline
$q$  &  power law $q^{ - 1.5}$  & Normal(10,3)\\
\hline
$N$  &  power law $N^{ - 3.4}$  & Normal(10,3)\\
\hline 
\end{tabular}
\end{center} 
\end{table}

\begin{figure}
\narrowtext
\begin{center}
\includegraphics[width=0.25\textwidth,angle=-90]{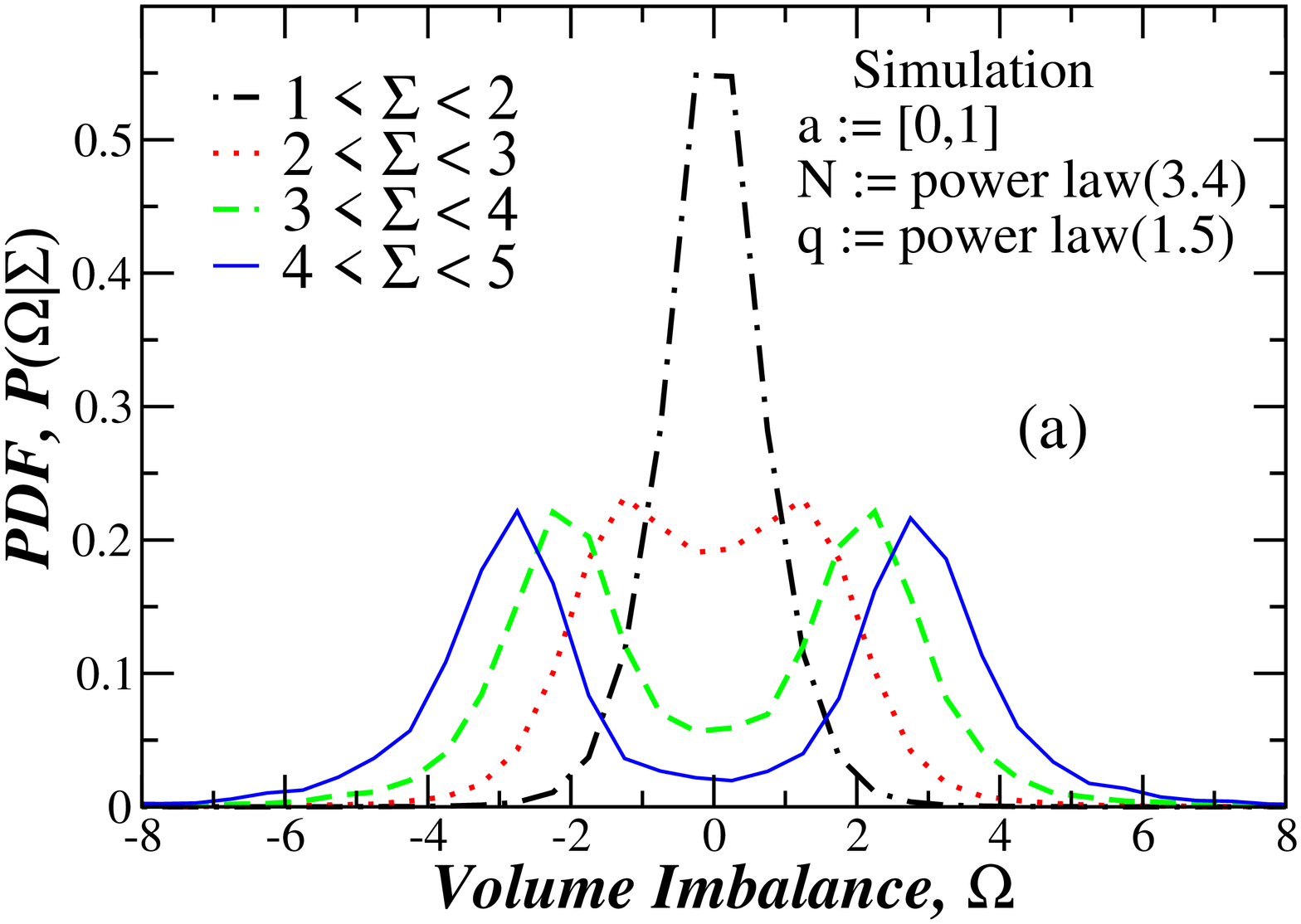}
\includegraphics[width=0.25\textwidth,angle=-90]{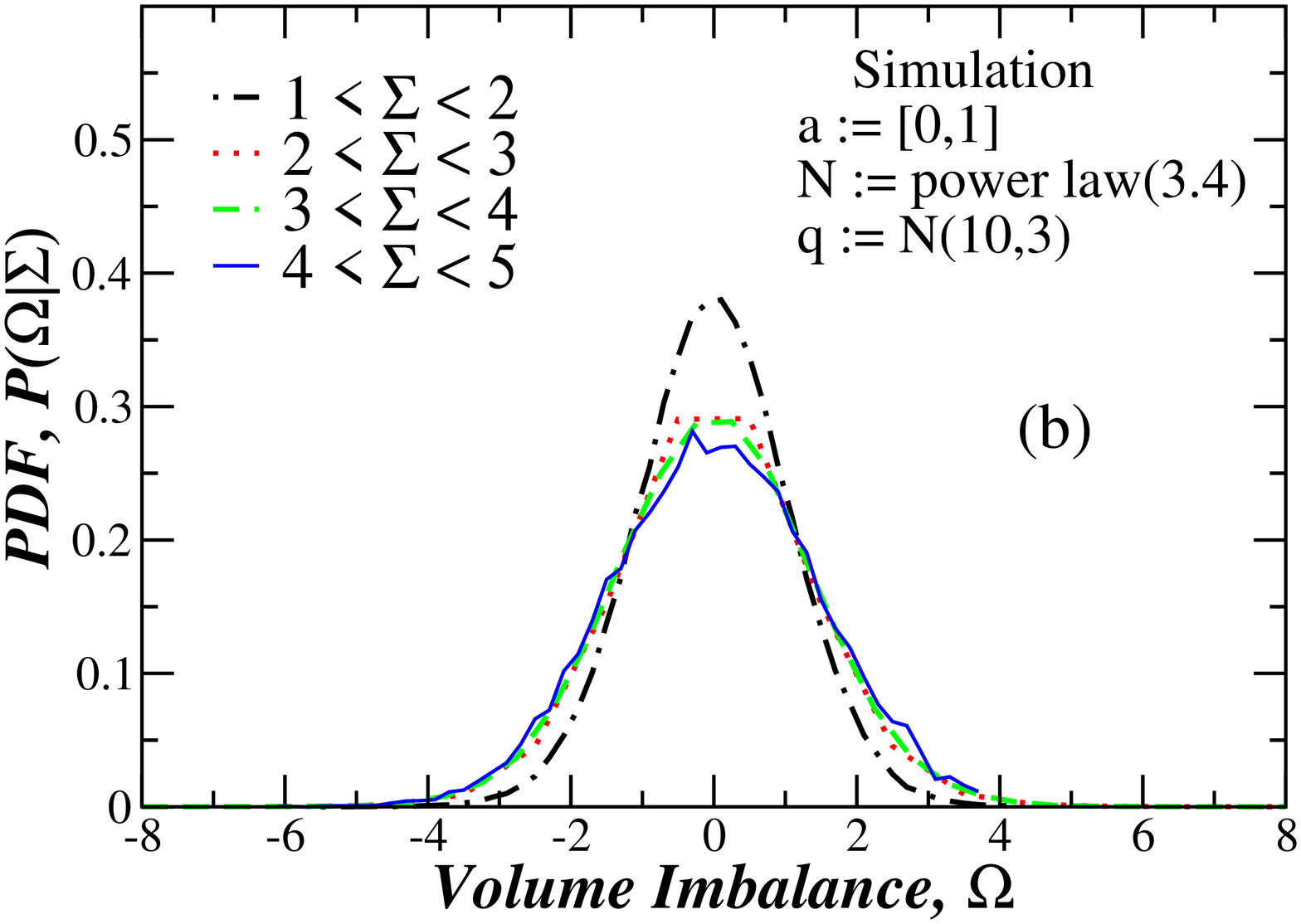}
\end{center}
\begin{center}
\includegraphics[width=0.25\textwidth,angle=-90]{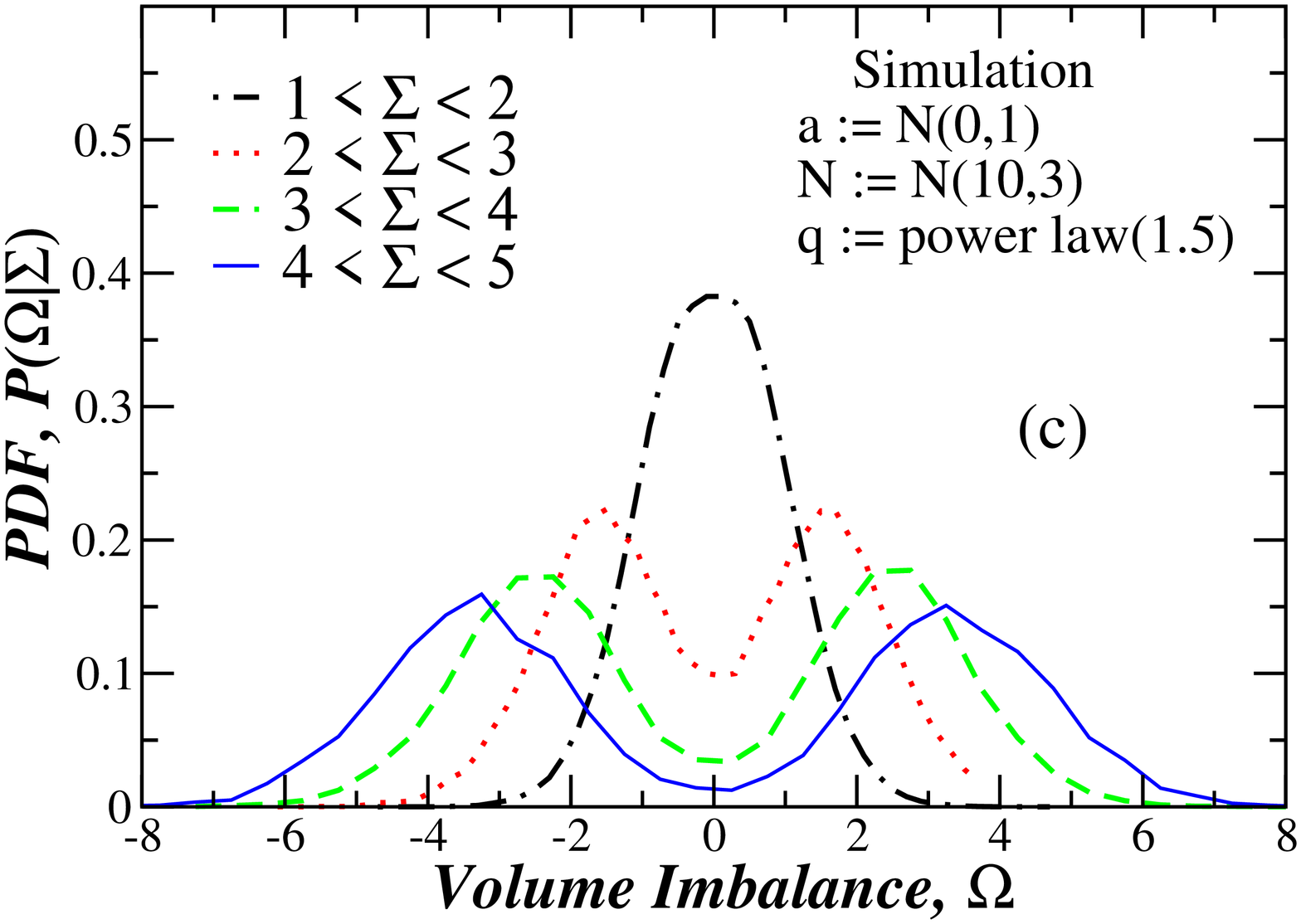}
\includegraphics[width=0.25\textwidth,angle=-90]{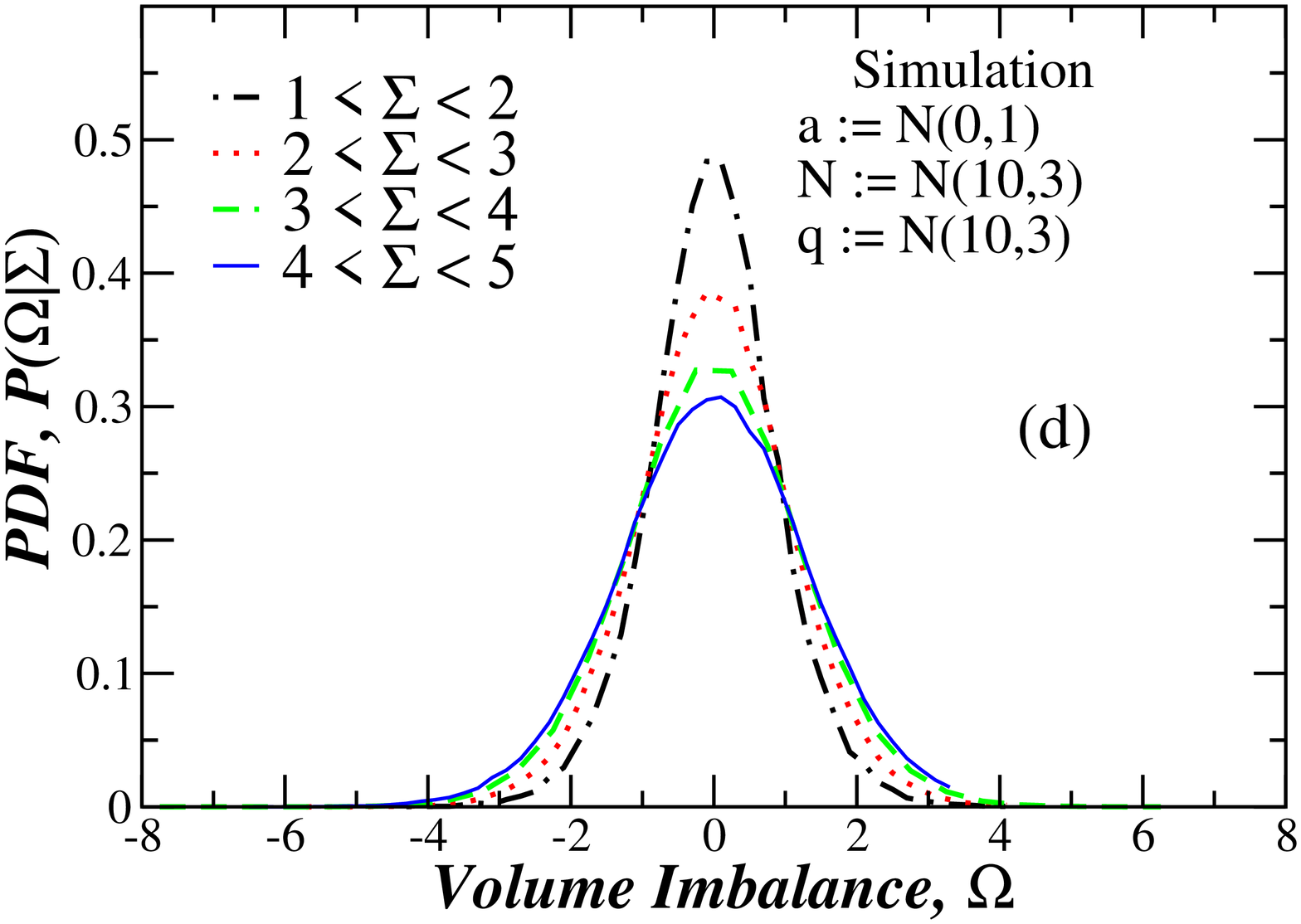}
\end{center}
\caption{ $P(\Omega|\Sigma)$ with $\Omega$ and $\Sigma$ generated from
i.i.d., for four different combinations of $a,~q,~N$. We observe a
bi-modal distribution of $P(\Omega|\Sigma)$ only when $P(q)$ has a power
law functional form. Since $P(q) \sim q^{-1.5}$ and the definition of
$\Omega$ has $q$ in its argument, $P(\Omega)$ also has the functional
form of a power law. Thus for the power law distributed $P(\Omega)$
occurrence of extreme positive and negative events are much more
probable than if $P(\Omega)$ did not have a fat tail. $\Sigma$ by
definition is the standard deviation of the random variable
$\Omega$. Large values of $\Sigma$ occurs when extreme events (both
positive and negative) are sampled. Thus $P(\Omega|\Sigma)$ has two
extremas resulting from a large positive and negative sampling of the
random variable $\Omega$ when groups with large $\Sigma$ values are
chosen [c.f. (a) and (c)].}
\label{sim}
\end{figure}

Since $a$, $q$, and $N$ are i.i.d, there are no autocorrelations like
that in the TAQ time series, but even then, in we find
$P(\Omega|\Sigma)$ to have a bi-modal distribution only when the PDF of
the share volume is a power law (c.f. Fig 3). This observation can be
explained as follows: First note that the PDF $P(q)$ of $q$ has the
functional form of a power law, i.e., $P(q) \sim q^{-1.5}$. Next note
that the $\Omega$ has a $q$ in its definition, thus the PDF $P(\Omega)$ of
$\Omega$ also has the functional form of a power law. Occurrence of
extreme positive and negative events are much more probable when
$P(\Omega)$ has a fat tail. Since $\Sigma$ by definition is the standard
deviation of the random variable $\Omega$, large values of $\Sigma$
occur when extreme events (both positive and negative) of $\Omega$ are
sampled. Thus $P(\Omega|\Sigma)$ has two extrema resulting from large
positive and negative sampling of the random variable $\Omega$ when we
choose groups with large $\Sigma$ values [c.f. Fig.~\ref{sim}(a) and
~\ref{sim}(c)].

Table~\ref{pcor} tabulates the estimated correlations between the pairs
$q$, $\Sigma$, and $N$, $\Sigma$ for the case of TAQ database and
simulation. We find correlation values for the TAQ database and
simulation to be statistically similar when $q$ and $N$ are power law
distributed, which further demonstrates that the observed two phase
effect is {\bf not} a signature of hidden collective phenomena within
financial market. An explanation similar to the one above is reported in
ref.~\cite{Bouchaud}.

\begin{table}
\begin{center}
\caption{Correlations of pair $(q, \Sigma)$ and $(N, \Sigma)$.}
\label{pcor}
\begin{tabular}{|c|c|c|}
\hline
Correlation                  & $(q,\Sigma)$ & $(N,\Sigma)$ \\
\hline
TAQ 2000-2001 database       & 0.69       & 0.01 \\
\hline
Simulation  with $q$, $N$    &            &      \\
distributed as in Fig 3a     & 0.95       & 0.00 \\
\hline
Simulation  with $q$, $N$    &            &      \\
distributed as in Fig 3b     & 0.08       & 0.03 \\
\hline
Simulation  with $q$, $N$    &            &      \\
distributed as in Fig 3c     & 0.71       & 0.00 \\
\hline
Simulation  with $q$, $N$    &            &      \\
distributed as in Fig 3d     & 0.19       & 0.11 \\
\hline
\end{tabular}
\end{center}
\end{table}

\begin{figure}
\begin{center}
\includegraphics[width=0.32\textwidth,angle=-90]{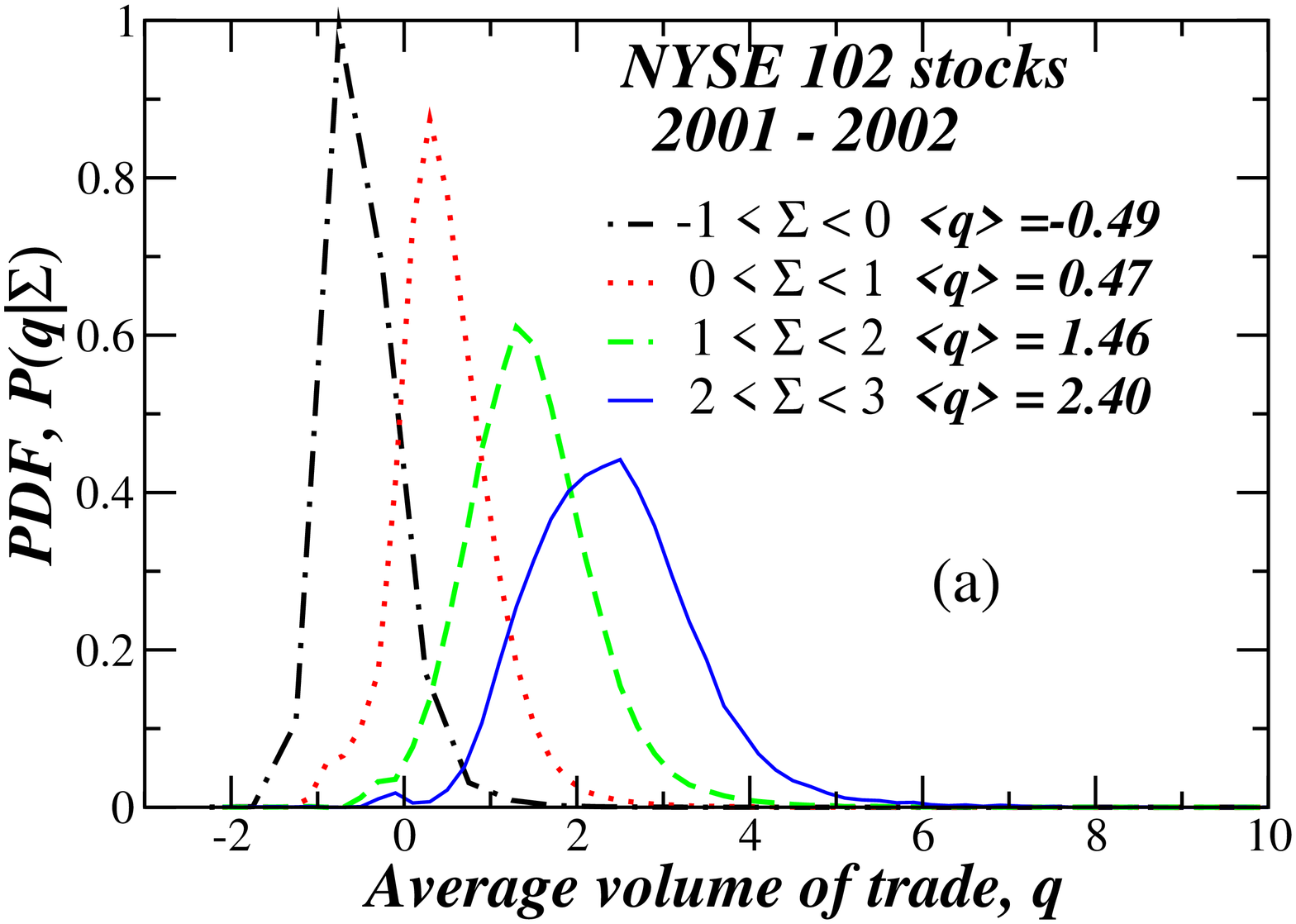}
\end{center}
\begin{center}
\includegraphics[width=0.32\textwidth,angle=-90]{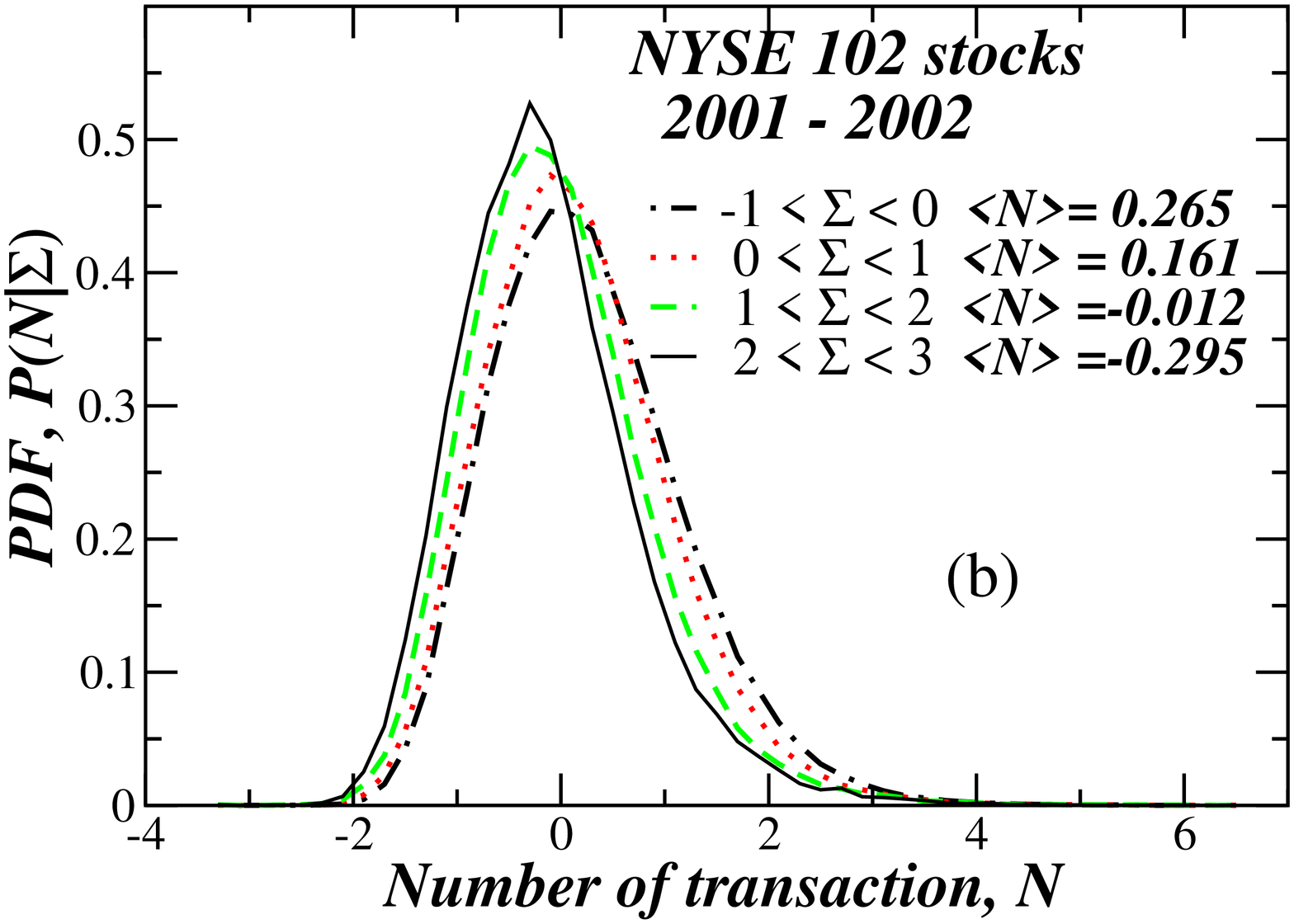}
\end{center}
\caption{(a) Probability density function $P(q|\Sigma)$ of the volume of
shares for given $\Sigma$. Note that $q$, $N$, and $\Sigma$ are
normalized quantities with their average value subtracted. We observe
that the transaction volumes are higher for groups with large
$\Sigma$. (b) PDF $P(N|\Sigma)$ of $N$ for given $\Sigma$. We observe
that the mean number of transaction $<N>$ slightly decreases for groups
with high $\Sigma$.}
\label{Sigma}
\end{figure}

To examine this in closest detail, we next estimate the PDF of $q$ and
$N$ for a given $\Sigma$ from the TAQ database. Fig.\ref{Sigma}(a) show,
that $q$ is large for large $\Sigma$. Fig.\ref{Sigma}(b) shows that $N$
is almost independent of $\Sigma$. By this, we can infer that the
bi-modal PDF $P(\Omega|\Sigma)$ is caused by the transactions of large
investors with large number of shares. The small decrease in $<N>$ as
$\Sigma$ increases has a simple explanation, which was pointed out by
our referee. In normal markets the specialist, whose role is to maintain
a fair and orderly market, crosses the book via his clerk very regularly
in an almost automated fashion. When fluctuations become large the
specialist must take a close look at the electronic order book and at
the order from the floor before he may decide which and at what price it
is fair to cross the orders. This manual intervention takes time and is
a plausible cause of the decrease in the number of trades.

\begin{figure}
\narrowtext
\begin{center}
\includegraphics[width=0.25\textwidth,angle=-90]{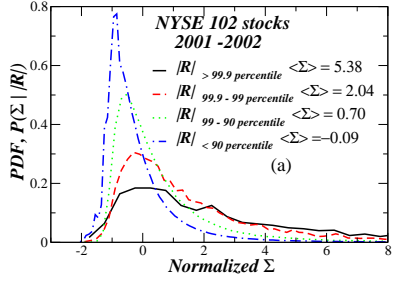}
\includegraphics[width=0.25\textwidth,angle=-90]{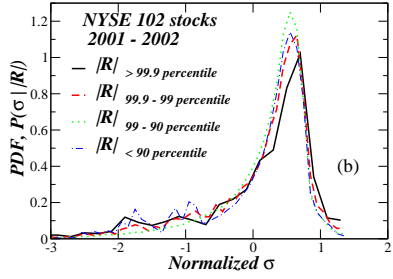}
\end{center}
\begin{center}
\includegraphics[width=0.25\textwidth,angle=-90]{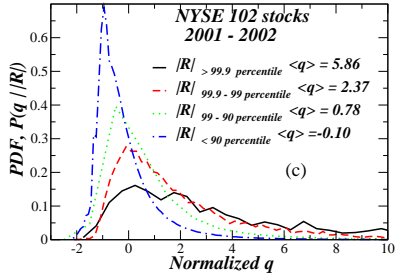}
\includegraphics[width=0.25\textwidth,angle=-90]{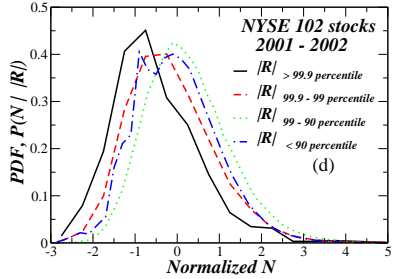}
\end{center}
\caption{(a) PDF $P(\Sigma||R|)$ of $\Sigma$ for a given $|R|$. We
observe that rarely occurring high $|R|$ results in higher volume
fluctuations. (b) PDF $P(\sigma||R|)$ of $\sigma$ for a given $|R|$. The
fluctuation of the number imbalance $\sigma$ is independent of $|R|$. (c)
PDF $P(q||R|)$ of $q$ for a given $|R|$. We observe that high $|R|$
prompts the average volume of shares per trade $q$ to increase. (d) PDF
$P(N||R|)$ of $N$ for a given $|R|$. We observe a slight decrease in the
number of transactions with an increase in $|R|$.}
\label{Return}
\end{figure}

To have a better understanding of how the absolute value of the price
fluctuations $|R|$ effects $\Sigma,~\sigma,~q,~N$, we next study the PDF
$P(\Sigma||R|),~P(\sigma||R|),~P(q||R|),~P(N||R|)$ for given $|R|$ using
the TAQ database. First we sort the database of $102 \times 26 \times
400$ items ( 102 stocks, 26 terms per a day, about 400 days in two
years) with respect to $|R|$. Next we divided these items into four
groups bordered by 99.9 percentile, 99 percentile and 90
percentile. Figure~\ref{Return} plots the PDF
$P(\Sigma||R|),~P(\sigma||R|),~P(q||R|),~P(N||R|)$ of each of these
groups.

We observe that even though $|R|$ is not correlated to $\Sigma$ or
$q$~[c.f table~\ref{corr}], the PDF $P(\Sigma||R|),~P(q||R|)$ exhibits
significant differences among different groups of $|R|$. This is because
rarely occurring large $|R|$ causes large fluctuations but does not
significantly contribute to correlation.

In contrast other quantities such as $\sigma$ and $N$ show there is not
so much of a difference. Fig.\ref{Return}(d) again shows the same effect
as seen in Fig.\ref{Sigma}(b) where large fluctuations decrease the
number of transactions within a specified time interval.

In conclusion it can be inferred that the ``two-phase behavior'' as
reported in~\cite{Stanley1} gives no evidence as to whether critical
phenomena exist in the financial market or not. The observed out of
equilibrium phase is a feature of the power law PDF of the share
volume. The absolute value of price fluctuation causes a large
fluctuation in the share volume, and large fluctuation of the absolute
price causes a decrease in the number of trades in a specified time
interval.

We thank S.~V.~Buldyrev, Y.~Lee for helpful discussions and suggestions
and K.~M thanks the NSF for financial support.


\end{document}